\input harvmac
\newcount\figno
\figno=0
\def\fig#1#2#3{
\par\begingroup\parindent=0pt\leftskip=1cm\rightskip=1cm\parindent=0pt
\global\advance\figno by 1
\midinsert
\epsfxsize=#3
\centerline{\epsfbox{#2}}
\vskip 12pt
{\bf Fig. \the\figno:} #1\par
\endinsert\endgroup\par
}
\def\figlabel#1{\xdef#1{\the\figno}}
\def\encadremath#1{\vbox{\hrule\hbox{\vrule\kern8pt\vbox{\kern8pt
\hbox{$\displaystyle #1$}\kern8pt}
\kern8pt\vrule}\hrule}}

\overfullrule=0pt

%
\def\underarrow#1{\vbox{\ialign{##\crcr$\hfil\displaystyle
 {#1}\hfil$\crcr\noalign{\kern1pt\nointerlineskip}$\longrightarrow$\crcr}}}
%

\def\bar{\overline}

\def\ap#1#2#3{Ann. Phys. {\bf #1} (#2) #3}

\def\inbar{\vrule height1.5ex width.4pt depth0pt}
\def\IC{\relax\hbox{\kern.25em$\inbar\kern-.3em{\rm C}$}}
\def\IR{\relax\hbox{\kern.25em$\inbar\kern-.3em{\rm R}$}}
\def\IZ{\relax\ifmmode\hbox{Z\kern-.4em Z}\else{Z\kern-.4em Z}\fi}

\font\zfont = cmss10 

\def\bigone{\hbox{1\kern -.23em {\rm l}}}
\def\ZZ{\hbox{\zfont Z\kern-.4emZ}}


\def\drawbox#1#2{\hrule height#2pt
        \hbox{\vrule width#2pt height#1pt \kern#1pt
              \vrule width#2pt}
              \hrule height#2pt}

\def\Asym#1#2{\vcenter{\vbox{\drawbox{#1}{#2}
              \kern-#2pt       
              \drawbox{#1}{#2}}}}

\batchmode
  \font\bbbfont=msbm10
\errorstopmode
\newif\ifamsf\amsftrue
\ifx\bbbfont\nullfont
  \amsffalse
\fi
\ifamsf
\def\IR{\hbox{\bbbfont R}}
\def\IC{\hbox{\bbbfont C}}

\def\IZ{\hbox{\bbbfont Z}}


\midinsert
\endinsert


\nref\uno{M.B. Green, J.H. Schwarz and E. Witten, {\it Superstring Theory} vols I and
II (Cambridge: Cambridge University Press 1987).}

\nref\dos{ J. Polchinski, {\it String Theory} vols I and II (Cambridge: Cambridge
University Press 1998).}
 
\nref\tres{D. L\"ust and S. Theisen, {\it Lectures on String Theory} (Berlin:
Springer-Verlag 1989).}

\nref\cua{J. Polchinski, ``TASI Lectures on D-Branes,'' hep-th/9611050.}

\nref\cinc{C.V. Johnson, ``D-Brane Primer,'' hep-th/0007170.}

\nref\seis{A. Vilenkin 1994,{\it Cosmic Strings and Other Topological Defects} (Cambridge:
Cambridge University Press 1994).} 

\nref\siete{ G. t' Hooft, ``Monopoles, Instantons and Confinement,'' hep-th/0010225.}

\nref\ocho{H. Kleinert, {\it Gauge Fields in Condensed Matter}, two vols. (World Scientific:
1989).}

\nref\trece{E. Witten, ``Superconducting Strings,'' Nucl. Phys. B {\bf 249} (1985) 557.}

\nref\diez{A. Sen, ``Non-BPS States and Branes in String Theory'', hep-th/9904207; J.H. Schwarz, ``TASI
Lectures on Non-BPS D-Brane Systems'', hep-th/9908144; A. Lerda and R. Russo, ``Stable non-BPS States in
String Theory, hep-th/9905006.} 

\nref\douglas{M.R. Douglas and N.A. Nekrasov, ``Noncommutative Field Theory'', hep-th/0106048.}

\nref\szabo{R.J. Szabo, ``Quantum Field Theory and Noncommutative Spaces'', hep-th/0109162.}

\nref\hklm{J. Harvey, P. Kraus, F. Larsen, and E.J. Martinec, ``D-branes and Strings as
Non-commutative Solitons,'' JHEP {\bf 0007} (2000) 042, hep-th/0005031.}

\nref\gorsky{A.S. Gorsky, Y.M. Makeenko and K.G. Selivanov, ``On Noncommutative Vacua and Noncommutative
Solitons'',  Phys. Lett. B {\bf 492} (2000) 344, hep-th/0007247.}

\nref\hklmdos{J. Harvey, P. Kraus and  F. Larsen, ``Exact Noncommutative Solitons'', JHEP {\bf 0012}
(2000) 024, hep-th/0010060.}

\nref\doce{R. Gopakumar, S. Minwalla, and A.
Strominger, ``Noncommutative Solitons,'' JHEP {\bf 0005} (2000) 020, hep-th/0003160.} 

\nref\dasgupta{K. Dasgupta, S. Mukhi and G. Rajesh, ``Noncommutative Tachyons'', JHEP {\bf 0006}
(2000) 022, hep-th/0005006.}

\nref\witten{E. Witten, ``Noncommutative Tachyons and String Field Theory'', hep-th/0006071}

\nref\sochi{C. Sochichiu, ``Noncommutative Tachyon Solitons, Interaction with Gauge Fields'',
JHEP {\bf 08} (2000) 026, hep-th/0007217.}

\nref\gopa{R. Gopakumar, S. Minwalla and A. Strominger, ``Symmetry Restoration and Tachyon
Condensation in Open String Theory'', hep-th/0007226.}

\nref\seiberg{N. Seiberg, ``A Note on Background Independence in Noncommutative Gauge Theories,
Matrix Model and Tachyon Condensation'', JHEP {\bf 09} (2000) 003, hep-th/0008013.}

\nref\asen{A. Sen, ``Some Issues in Noncommutative Tachyon Condensation'', hep-th/0009038.}

\nref\wittenk{E. Witten, ``Overview of K-Theory Applied to Strings'', hep-th/0007175.}

\nref\harmoore{J.A. Harvey and G. Moore, ``Noncommutative Tachyons and K-Theory'', hep-th/0009030.}

\nref\marmoore{E.J. Martinec and G. Moore, ``Noncommutative Tachyons on Orbifolds'', hep-th/0101199.}

\nref\harvey{J.A. Harvey, ``Komaba Lectures on Noncommutative Solitons and D-branes'', hep-th/0102076.}

\nref\tatar{R. Tatar, ``A Note on Noncommutative Field Theory and Stability of Brane-Antibrane
Systems'', hep-th/0009213.}

\nref\raja{P. Kraus, A. Rajaraman and  S. Shenker, ``Tachyon Condensation in Noncommutative Gauge
Theory'',  Nucl. Phys. B {\bf 598} (2001) 169, hep-th/0010016.}

\nref\finn{J.A. Harvey, P.Kraus and F. Larsen, ``Exact Noncommutative Solitons'', JHEP {\bf 0012}
(2000) 024, hep-th/0010060.}

\nref\das{K. Dasgupta and Z. Yin, ``Nonabelian Geometry'', hep-th/0011034.}

\nref\zachos{C.M. Zachos, ``Deformation Quantization: Quantum Mechanics Lives and Works in
Phase-Space'', hep-th/0110114.} 

\nref\dseis{S. de Alwis, ``Tachyon Condensation in Rotated Brane
Configurations,'' Phys. Lett. B {\bf 461} (1999) 329, hep-th/9905080.}
 
\nref\vuno{N. Seiberg and E. Witten, ``String Theory and Noncommutative
Geometry,'' JHEP {\bf 09} (1999) 032, hep-th/9908142.}

\nref\veinte{V. Schomerus, ``D-branes and Deformation Quantization,'' JHEP {\bf 9906}
(1999) 030, hep-th/9903205.}

\nref\vtres{H. Garc\'{\i}a-Compe\'an, J.F. Pleba\'nski, M. Przanowski, and F.J. Turrubiates,
``Deformation Quantization of Bosonic Strings,'' J. Phys. A {\bf 33} (2000) 7935, hep-th/0002212.}

\nref\vcuatro{L. Pilo and A. Riotto, ``The Non-commutative Brane World,'' JHEP {\bf 03} (2001) 015, 
hep-ph/0012174.}


\Title{hep-th/0110119, CINVESTAV-FIS-01/077}
{\vbox{\centerline{ Remarks on Noncommutative Solitons}}}
\medskip
\smallskip
\centerline{Hugo
Garc\'{\i}a-Compe\'an\foot{E-mail address: {\tt compean@fis.cinvestav.mx}} and Jorge
Moreno\foot{E-mail address: {\tt jmoreno@fis.cinvestav.mx}}}
\smallskip
\centerline{\it Departamento de F\'{\i}sica}
\centerline{\it Centro de Investigaci\'on y de Estudios Avanzados del IPN}
\centerline{\it Apdo. Postal 14-740, 07000, M\'exico D.F., M\'exico}

\bigskip 
\medskip 
\vskip 1truecm 
\noindent 
In the first part of this work we consider an unstable non-BPS $Dp-\bar{Dp}$-brane pair in Type II
superstring theory. Turning on a background NS-NS $B$-field (constant and nonzero along two
spatial directions),
we show that the tachyon responsible for the unstability has a complex GMS solitonic solution, which is
interpreted as the low energy remnant of the resulting $D(p-2)$-brane. In the second part, we apply these
results to construct the noncommutative soliton analogous of Witten's superconducting string. This
is done by considering the
complex GMS soliton arising from the $D3-\bar{D3}$-brane annihilation in Type IIB superstring theory. In
the presence of left-handed fermions, we apply the Weyl-Wigner-Moyal correspondence and the bosonization
technique to show that this object behaves like a superconducting wire.


\noindent

\Date{October, 2001}

\newsec{Introduction}

Recent breakthroughs in string theory have not only proven to be both insightful
and breathtaking, but have rapidly overturned obsolete notions initially thought
to be well established \refs{\uno,\dos,\tres}. Particularly, the discovery of $D$-branes
in the nonperturbative regime has revealed a deeper underlying structure, which
might be a first glance of an ultimate theory \refs{\dos,\cua,\cinc}.

On the other hand, topological defects (see for instance, Refs. \refs{\seis,\siete}) in field theory
have been studied
for a number of years. The traditional approach was to consider these as a consequence
of the spontaneous breakdown of gauge symmetries. The spirit then was to explore
the nonperturbative sector of the Standard Model and Grand Unified Theories containing large Lie groups.
Also, a plethora of defects, ranging from monopoles and vortices to kinks and domain
walls, have been obtained in both particle physics and condensed matter systems \refs{\ocho}. 
It is important to mention a very interesting example: the {\it superconducting string} found by Witten in
Ref. \trece. The idea there is to consider a four-dimensional scalar theory with a $U(1) \otimes
\widetilde{U}(1)$ gauge symmetry. The spontaneous breakdown of one of the $U(1)$'s yields a
string-like
solution; while the breaking of the remnant $U(1)$ in the core of the string endows this object with
superconductivity. In conclusion:
if gauge symmetries are truly present in nature, such defects ought to exist
and they should be found experimentally. 

A more modern application of topological defects is in the understanding of
$Dp$-brane anti-$Dp$-brane annihilations. Such configurations are
non-BPS (for a review, see for instance Ref. \refs{\diez}), and they are unstable due to the presence of
a
tachyonic mode
in their worldvolume. By finding a suitable vortex-like configuration for the
tachyon field, it has been shown that the result of the above process is the
emergence of a stable BPS $D(p-2)$-brane.

Another outstanding new trend is the study of $D$-branes in the presence
of a NS-NS constant $B$- field. In the low energy
limit, its effect is the appearance of a Moyal $*$-product
in the fields participating in the Operator Product Expansion (OPE), thence
obtaining a noncommutative effective field theory (for a review, see \refs{\douglas,\szabo}).
A relevant feature here is that we can associate fields in $\IR^2_*$
to operators in the Hilbert space of a simple harmonic oscillator (SHO). This
association is known as the Weyl-Wigner-Moyal (WWM) correspondence.

Last year, Harvey, Kraus, Larsen and Martinec (HKLM) used this approach
to study the decay of a bosonic $D25$-brane into a $D23$-brane \refs{\hklm}.
Considerations of the classical vacua and its implications were considered 
in Ref. \refs{\gorsky}. Finite noncommutative parameter corrections were
performed in \refs{\hklmdos}. Large non-commutativity parameter approximation is simpler than that for
a finite parameter. In this paper we focus mainly in this approximation.
Bosonic string theory has a tachyon mode, which makes $D$-branes of all dimensions unstable \refs{\diez}.
However, instead of searching a tachyonic vortex configuration, HKLM found a nontrivial solution by
introducing a large $B$-field along two spatial directions in the $D25$-brane worldvolume. The solution is
the
real noncommutative soliton discovered by Gopakumar, Minwalla and Strominger (GMS) \refs{\doce}. This
object is identified with the remnant of the $D23$-brane. An extension of HKLM's work to Type II
superstring theory is
described in \refs{\dasgupta}. Further work on noncommutative tachyons in the large noncommutative
parameter approximation worked out in Refs. \refs{\witten,\sochi,\gopa,\seiberg,\asen}. A K-theoretic 
description of noncommutative tachyons is done in \refs{\wittenk,\harmoore}. The description of
tachyon condensation in orbifolds is discussed in Ref. \marmoore.
For a recent review on the subject, for instance, see Ref. \refs{\harvey}.

The objective of the first part of this work is to apply HKLM's idea \refs{\hklm} and study the case of
Type II superstring $Dp-\overline{Dp}$
annihilation in the presence of a large $B$-field along two spatial directions. Many techniques used by
GMS in Ref. \doce, such as that of identifying nontrivial solutions with projection operators, are applied
in here as well. However, we now have a charged tachyon field under the Chan-Paton gauge symmetry
$U(1)\otimes \widetilde{U}(1)$. The solution is a complex GMS soliton,
and it is regarded as the low-energy remnant $D(p-2)$-brane. $Dp-\overline{Dp}$ pairs with 
$B$-field and non-zero magnetic fluxes were previously considered in Ref. \tatar\ and further
explored in Refs. \refs{\raja,\finn}. The generalization to nonabelian fluxes was studied at
Ref. \das.

In the second part of this paper, we attempt to construct an object similar to Witten's superconducting
string in the context of Noncommutative Solitons in string theory. The idea is to consider a non-BPS
$D3-\bar{D3}$-brane pair in Type IIB superstring theory in the presence of a large and constant NS-NS
$B$-field turned on along only two worldvolume spatial directions. The fact that the tachyon is charged
under a $U(1)\otimes \widetilde{U}(1)$ Chan-Paton gauge symmetry is a tantalizing similarity to
Witten's original setup. Therefore, using the results obtained in the first part of this work, we
identify
the complex GMS solitonic solution to the tachyon with the low-energy remnant of the $D1$-brane (which is
itself the product of the $D3-\bar{D3}$-brane annihilation). We coin the term {\it
noncommutative D-string} (or {\it
noncommutative string}, for short) to describe this object. Like in Witten's {\it string}, there are
left-handed fermions, which naturally arise from the supersymmetric spectrum of the open string attached
to the $D1$-brane. We shall just consider the flat space case; so in the low-energy regime, these
fermions
live in a space ${\cal M}^{1+1} \otimes {\IR}^2_*$ with the complex GMS soliton as a background
field\foot{ In the general case ${\cal M}^{n+1}$ denotes a Minkowski space with one timelike and $n$
spacelike dimensions, while ${\IR}_{*}^{s}$ denotes an $s$-dimensional noncommutative space. }.
Integrating out the transverse noncommutative directions, this complex GMS soliton projects out most of
the fermionic modes, leaving behind a simple two-dimensional theory. Happily, such theory can be exactly
solved by the technique of bosonization, inspired by Witten's method for the case of the superconducting
string \refs{\trece}. Surprisingly, we find that the fermionic current along $z$-direction doesn't decay.
Hence, our noncommutative string's behavior is similar to a superconducting wire.

This paper consists in two parts. The first one explains the construction of a complex GMS solitons
from $Dp-\bar{Dp}$ brane annihilation. The second part applies this idea to show the existence of a
noncommutative version Witten's superconducting string in the context of superstring theory. In
section 2, we overview the basic properties of the unstable non-BPS $Dp-\overline{Dp}$-brane
configuration in Type II superstring theory. Also, we explain how noncommutativity arises from the
NS-NS $B$-field and introduce the WWM correspondence. In section 3, we turn on a $B$-field on the
$Dp-\overline{Dp}$ brane system and find a complex gauge-coupled GMS soliton, which we identify with
a BPS $D(p-2)$-brane. In section 4, we study the case where $p=3$, and construct the noncommutative
$D$-string. We couple the four-dimensional left-handed fermions (coming from the supersymmetric
spectrum of the open string attached to the $D1$-brane) in the low energy regime to the background
GMS soliton. Then, by integrating out the two noncommutative directions, we obtain a two-dimensional
theory along the $D$-string. At the end we show, by applying the bosonization technique, that this
object appears to be superconducting.

\vskip 2truecm
\newsec{Basic Tools}

The purpose of this section is to give the reader a brief overview of the tools and ideas
necessary to address the problem of the $Dp-\overline{Dp}$ brane configuration with a $B$-field
background.  It must be pointed out, however, that our aim is not to provide an extensive
review of
noncommutative field theories. For a more complete discussion, see
\refs{\douglas,\szabo,\harvey}. For a very recent review on Weyl-Wigner-Moyal deformation
quantization see Ref. \zachos.

\vskip 1truecm
\subsec{$Dp-\overline{Dp}$-Brane Annihilation}

To begin with, consider a pair of parallel $Dp$-branes in Type II theory, with $p$ even in the
Type IIA and odd in the Type IIB theory.
This system is stable and BPS, and has a $U(1)\otimes \widetilde{U}(1)$ Chan-Paton internal symmetry.
Roughly
speaking, we can turn  one of the $Dp$-branes
into a  $\overline{Dp}$-brane by rotating it an angle $\pi$ in the transverse
directions \refs{\dseis}. A consequence of this is the reversal
of the GSO projection, hence the occurrence of a tachyon along with the previously
cancelled massive states. Thus, the $Dp-\overline{Dp}$-brane configuration obtained by
rotating one of them is no longer BPS.

In general, the presence of a tachyon is a signal of instability.
Under certain circumstances, such unstable non-BPS
systems may decay into stable BPS $D$-branes of lower dimensions. In the
case of  $Dp-\overline{Dp}$-brane annihilation, this system
may decay into a stable $D(p-2)$-brane \refs{\diez}.

The tachyon
in the $Dp-\overline{Dp}$-brane worldvolume is charged $(-1,+1)$
under the gauge symmetry $U(1)\otimes \widetilde{U}(1)$. Therefore,
the tachyonic lagrangian ${\cal L}_t$ is given by
\eqn\tachyon{
{\cal L}_{t}= \overline{D_{\mu }T}D^{\mu }T-V(T),}
where the covariant derivative is

\eqn\mct{
D_{\mu}T=(\partial _{\mu }-iA_{\mu }+i\widetilde{A_{\mu}}) T,}
while $A_{\mu }$ and $\widetilde{A_{\mu }}$ are real functions and they are respectively the gauge
fields of $U(1)$ and $\widetilde{U}(1).$  

The traditional method to find a stable $D(p-2)$-brane is as follows. First, we parametrize
the original $(p+1)$-dimensional worldvolume by the coordinates $(r,\theta , \widetilde{x}^a)$,
where $\widetilde{x}^a$ are longitudinal spacetime coordinates to the $D(p-2)$-brane. One must find a
cylindrically
symmetric vortex
configuration
localized in the vicinity of $r=0$ for the tachyon \refs{\diez}. Such a configuration is
required
to described a pure vacuum for large $r$ in a topologically nontrivial way.
This is achieved by imposing the following asymptotic behavior $(r\rightarrow \infty)$:

\eqn\vortex{ 
T\sim T_{min}e^{i\theta }, \ \ \  \  
A_{\theta}-\widetilde{A_{\theta }}\sim 1,}
where $T_{min}$ is the value in which $ V( T)  $ is minimized.

These conditions \vortex\ ensure that for large $r,$ $D_{\mu }T \to 0$
and $V(T) \to V( T_{min}),$
leaving a soliton placed in the small $r$ region. Notice that this soliton is
independent of $\widetilde{x}^a$. This is a vortex string\foot{
A similar situation is studied in \refs{\trece}. However, in that work $U(1)$
is spontaneously broken to give rise to the string and the other fields in $\widetilde{U}(1)$
make the string superconducting.}, and we identify it with a stable BPS $D(p-2)$-brane.

Nevertheless, imposing vortex-like asymptotic conditions as in \vortex\
is not the only method of obtaining stable nontrivial solutions of the tachyon
field. A few months back, it was found 
that the tachyon allows a different
type of solutions if some directions are noncommutative \refs{\doce}.

\vskip 1truecm
\subsec{Noncommutativity from String Theory: the $B$-field }

In string theory, the conventional low energy limit is to take $\alpha '\to 0$. The
result of this is the inevitable decoupling of the massive modes from the effective
theory. 

If we additionally turn on a constant NS-NS $B$-field, we still decouple
the massive modes from the theory. However, it turns out that one obtains a
nonlocal deformation of the field theory due to noncommutativity. This is a
stringy effect which helps display D$p$-branes as noncommutative solitons. 

Recall that Type II theories have a massless NS-NS symmetric background field $g_{\mu \nu }$ with $\mu,
\nu
=0,1,\cdots 9$, which we shall interpret as the background metric.  Likewise, these theories
contain an
antisymmetric field $B_{\mu \nu}$ in the massless NS-NS spectrum. These theories also admit R-R charged 
$Dp$-branes with open
strings attached to them. In this case, the OPE is merely 
$e^{ik_2\cdot X}e^{ik_1\cdot X}\sim (\tau -\tau ')^{2\alpha 'g^{\mu \nu} k_{1\mu
}\cdot k_{2\nu}}\times e^{i( k_{1}+k_{2}) \cdot X}+\cdots .$

Turning on this $B$-field, the OPE becomes $e^{ik_{1}\cdot X}(\tau) e^{ik_{2}\cdot X}( \tau ')$ $
\sim (\tau -\tau ') ^{2\alpha 'G^{\mu \nu }k_{1\mu }k_{2\nu}}$ $\times [ e^{-{i \over 2}\Theta
^{\mu \nu }k_{1\mu }k_{2\nu }}] \times e^{i( k_{1}+k_{2}) \cdot X}+\cdots ,$ where $G^{\mu \nu }=(
{1 \over g+2\pi \alpha 'B}g {1 \over g-2\pi \alpha 'B}) ^{\mu \nu }$ is the effective metric seen
by the open string modes, and $\Theta ^{\mu \nu }=-( 2\pi \alpha ') ^{2}( {1 \over g+2\pi \alpha
' B}B {1 \over g-2\pi \alpha ' B}) ^{\mu \nu }$ is known as the noncommutativity parameter matrix.

The above calculations were carried over by Seiberg and Witten in Ref. 
\refs{\vuno} (see also references therein), but the noncommutativity
interpretation was first given by Schomerus \refs{\veinte}.

The configuration space counterpart of the extra factor appearing in the OPE, {\it i.e.} $e^{{i \over
2}\Theta ^{\mu \nu }\partial _{\mu }\partial '_{\nu }}$, has a rather peculiar interpretation. This factor
gives rise precisely to the Moyal $*$-product (a not commutative, but still associative product), which has
been studied for a number of years as a key feature in an alternative description of quantum mechanics
known as Deformation Quantization.  Recently, this description was applied to the quantization of bosonic
strings \refs{\vtres}.

The presence of the Moyal $*$-product in the OPE means that fields in the
effective theory get multiplied as follows
\eqn\equis{
( f*g) (x) =  f(x)e^{{i \over 2}\Theta^{\mu \nu} \buildrel{\leftarrow} \over{\partial}_{\mu}
\buildrel{\rightarrow} \over {{\partial}'}_{\nu}} g( x') |_{x=x'}\neq (g*f)(x),}
whereas in the absence of $B$-field, they were simply multiplied
as $( f\cdot g) ( x) =f( x) g( x) =g(x)f(x)=(g\cdot f)(x).$
 
In conclusion, we can fix our $B$-field in any convenient way to obtain
a desired effective theory with the characteristic that along those directions
where $B\neq 0$, the worldvolume of the $D$-brane is noncommutative and fields are $*$-multiplied.

\vskip 1truecm
\subsec{The Weyl-Wigner-Moyal Correspondence}

The simplest configuration is when the constant $B$-field is nonzero only
along two spatial directions. Let's choose these to be $x$ and $y$ and call
the noncommutative $x-y$ plane $\IR^{2}_{*} $. Therefore

\eqn\rdos{
B_{\mu \nu} = \pmatrix{0 & B & \cdots  & 0 \cr
-B & 0 &  & \vdots \cr
\vdots  &  & \ddots  & \vdots\cr
0 & \cdots  & \cdots  & 0\cr},}
where $B=B_{12}=-B_{21}$. As a result, we obtain noncommutativity along
the $x-y$ plane:

\eqn\thetam{
[x,y] _{*}=i\Theta ,}
where $\Theta =\Theta^{12}=-\Theta^{21}$ is the noncommutativity parameter and $[x,y]_* \equiv 
x*y -y * x$ is the Moyal bracket. 

Rescaling the coordinates to $x\to {x \over \sqrt{\Theta }}$ and $y\to {y\over \sqrt{\Theta }},$
we obtain the following commutator:
\eqn\rcom{
[ x,y] _{*}=i,}
which is very similar to $[ \widehat{q},\widehat{p}] =i,$
the position and momentum commutator of a quantum particle in one spatial dimension.
With this identification, calculations along noncommutating directions are straightforward. 

As in deformation quantization, we can associate fields $f(x,y)$ in the
noncommutative plane $\IR_{*}^{2}$ to operators $\widehat{{\cal O}_{f}}(
\widehat{q},\widehat{p})$ in the quantum particle's Hilbert space ${\cal H}
(\widehat{q},\widehat{p})$. The common identification is performed by using the WWM
prescription \refs{\harvey,\zachos}:

\eqn\weyl{
\widehat{O_{f}}( \widehat{q},\widehat{p}) = {1 \over ( 2\pi ) ^{2}}\int
dk_{q}dk_{p}\widetilde{f}( k_{q},k_{p}) \widehat{U}(
\widehat{q},\widehat{p}),}
where  $\widehat{U}( \widehat{q},\widehat{p}) =e^{-i(
k_{q}\widehat{q}+k_{p}\widehat{p}) }$ is a unitary operator, and the Fourier transform is just
$\widetilde{f}( k_{q},k_{p}) =\int dqdpf( q,p) e^{i( k_{q}q+k_{p}p) }.$
Therefore, we can write the operator
\eqn\wwm{
\widehat{O_{f}}( \widehat{q},\widehat{p}) = {1 \over ( 2\pi ) ^{2}}\int
dqdpf( q,p) \widehat{\Omega }( \widehat{q},\widehat{p}),}
where

\eqn\quantizer{
\widehat{\Omega }( \widehat{q},\widehat{p}) =\int dk_{q}dk_{p}e^{i(
k_{q}q+k_{p}p) }\widehat{U}( \widehat{q},\widehat{p})}
is known as the Stratonovich-Weyl quantizer \refs{\vtres}. 

A major consequence of this correspondence is that now it is easier to perform
integrations along $\IR_{*}^{2}$. Thus, with the above prescription,
it can be shown that

\eqn\trace{
{1 \over 2\pi \Theta }\int dqdp \ f(q,p) =Tr_{\cal H}
\bigg(\widehat{O_{f}}(\widehat{p},\widehat{q}) \bigg).}
Furthermore, another property is that in general, for any complex field $\vartheta$ living
in $\IR_{*}^{2}$,

\eqn\cuadratico{
\int dx dy \  \overline{\vartheta }*\vartheta =\int dx dy \  \overline{\vartheta }\vartheta.}

There are other convenient ways to work with fields in a noncommutative space.
Let's parametrize $\IR_{*}^{2}$ with complex coordinates $w= {1 \over \sqrt{2}}( x+iy)$ 
and $\overline{w}= {1 \over \sqrt{2}}( x-iy)$
and rescale them, so we are left with the following commutator:
\eqn\wcom{
[ w,\overline{w}] _{*}=1.}

Notice that this is the analogous to the quantum SHO commutator: 
$[ \widehat{a},\widehat{a}^{\dagger }]
=1,$ where $\widehat{a}$ is the annihilation operator and $\widehat{a}^{\dagger}$ the creation
operator.

The above results can be easily generalized to the case where there are $n$ pairs of noncommuting
coordinates. In general, a field $\varphi$ in ${\cal G}^{(p-2n)+1}\otimes \IR_{*}^{2n}$
can be expressed as\foot{ $x^a$ live in ${\cal G}^{q+1}$ which is a manifold that reduces to the
Minkowski space ${\cal M}^{q+1}$ as the metric $G_{ab}$ goes to a flat metric 
$\eta _{ab}$. Likewise, $x^i$ live in a $2n$-dimensional noncommutative space $\IR_{*}^{2n}=
\IR_{*}^{2}\oplus \cdots \oplus
\IR_{*}^{2}$ ($n$-times).}  

\eqn\expand{
 \varphi(x^{\mu }) =\sum _{m,n}\varphi _{mn}(
x^{a}) \Phi _{mn}( x^{i}),}
where the $ \Phi _{mn}(x^{i})  $ in $\IR_{*}^{2n}$ are related to $ \left|
m\right\rangle \left\langle n\right| $ in ${\cal H}_{n}=
{\cal H}(\widehat{q}_{1},\widehat{p}_{1})\oplus \cdots \oplus 
{\cal H}(\widehat{q}_{n},\widehat{p}_{n})$ \refs{\vcuatro}.

Further generalizations are overwhelmingly challenging, and beyond the scope
of this work.

\vskip 2truecm
\newsec{The $D(p-2)$-Brane as a Noncommutative Soliton}

In Ref. \refs{\hklm}, HLKM studied a process where a bosonic $D25$-brane decays into a $D23$-brane in
the presence of a large $B$-field. They found a nontrivial solution to the real
tachyon in the $D25$-brane. It was precisely the real GMS soliton \refs{\doce} which they
identified with the remnant of the $D23$-brane in the low energy limit. In this section,
we will apply this idea to the complex gauge-coupled tachyon in the Type II $Dp-\overline{Dp}$ -brane
system with a constant and large background $B$-field \refs{\harvey}.

\vskip 1truecm
\subsec{$Dp-\overline{Dp}$-Brane annihilation in the presence
of a $B$-field}

Recall that the non-BPS $Dp-\overline{Dp}$ configuration is unstable
because of the presence of a tachyon in its $(p+1)$-dimensional worldvolume. This tachyonic field
has charge $ +1 $ under the group $ U(1) $ with gauge fields $A_{\mu }$ and charge 
$-1$ under $\widetilde{U}(1)$ with gauge fields $\widetilde{A_{\mu }}$.
Consider a constant background $B$-field of the form given in \rdos, so the worldvolume
is $\Sigma _{p+1}={\cal G}^{(p-2)+1}\otimes \IR_{*}^{2}$. We will just focus on the case when
the worldvolume metric is flat $G_{\mu \nu }= \eta _{\mu \nu },$
thus $\Sigma_{p+1}={\cal M}^{(p-2)+1}\otimes \IR_{*}^{2}.$

Parametrizing the $x-y$ plane with the complex variables $( w,\overline{w})$, 
and the commutative coordinates $\widetilde{x}^a$, the $(p+1)$-dimensional action
is

\eqn\tstar{
S_{t}^{(p+1) }=\int_{\Sigma_{p+1}} d^{p-1}\widetilde{x} d^{2}w \bigg( \overline{D^{\mu }T}*D_{\mu
}T-V_{*}(
T,\overline{T})  \bigg),}
where the covariant derivative is $D_{\mu }T=\partial _{\mu }T-iA_{\mu }*T+i\widetilde{A_{\mu }}*T.$

Denoting $R_{\mu }=A_{\mu }-\widetilde{A_{\mu }}$, we are left with

\eqn\scuatror{
S_{t}^{(p+1)}=\int_{\Sigma_{p+1}} d^{p-1} \widetilde{x} d^{2}w \bigg( ( \partial ^{\mu
}\overline{T}+iR^{\mu
}*\overline{T}) *( \partial _{\mu }T-iR_{\mu }*T) -V_{*}(
T,\overline{T}) \bigg).}

Our considerations apply to generic potentials $V(T,\bar{T})$, but we will, for definiteness, mostly
discuss those of polynomial form: 

\eqn\poly{
V_{*}( T,\overline{T}) =V_{*}( \overline{T}*T) =\sum ^{n}_{k=1}a_{k}(
\overline{T}*T) ^{k},}
where, of course, we have abbreviated

\eqn\rlongdos{ (\bar{T} * T)^k = (\bar{T} * T)* (\bar{T} * T)* \dots * (\bar{T} * T), \ \ \ \ \
(k-{\rm times})}
and $k$ is a positive integer.

Also, as in Ref. \doce, in order for nontrivial solutions to exist, our potential $V_*(T,\bar{T})$ must
have at least two local minima. 

Let's now proceed and construct a simple solution. Recall that, in the absence of noncommutativity, we
found a vortex solution \vortex\ independent of $\widetilde{x}^a$. In the next section, we will be
interested in a solution with the same spacetime dependence (as in the vortex case):

\eqn\trdos{
T=T( w,\overline{w}),\ \ \ \ \ \  \overline{T}=\overline{T}( w,\overline{w}).}

The action is given by

\eqn\scuatrotheta{
S_{t}^{(p+1) }=\int_{\Sigma_{p+1}}d^{p-1} \widetilde{x} d^{2}w\bigg( - {1 \over \Theta }\partial
^{\overline{w}}\overline{T}*\partial _{w}T+ {i \over \sqrt{\Theta
}}(R^{\mu }*\overline{T}*\partial _{\mu }T-\partial ^{\mu }\overline{T}*R_{\mu }*T)+R^{\mu
}*\overline{T}*R_{\mu }*T-V_{*} \bigg).}

From now on, we will focus exclusively on the case of infinite noncommutativity, $\Theta \to \infty,$
since in this limit, the kinetic term along the noncommutative plane vanishes. We could eventually
introduce finite-$\Theta$ kinetic contributions in terms of a ${1 \over \Theta}$-expansion. However,
for
the time being, we want to keep the theory simple and focus more on the properties of the solitonic
solutions that depend on the potential $V_*(\bar{T},T)$.  Thus, we are left with

\eqn\scuatroeff{
S_{t}^{(p+1) }=\int_{{\Sigma}_{p+1}}d^{p-1} \widetilde{x} d^{2}w \bigg( R^{\mu
}*\overline{T}*R_{\mu }*T-V_{*}(\overline{T}*T) \bigg).}

Defining the potential

\eqn\veff{
\widetilde{V_{*}}( \overline{T},T) =\overline{R^{\mu }T}R_{\mu }T-V_{*}(
\overline{T}*T),}
the following simple action in ${\cal M}^{(p-2)+1}\otimes \IR_{*}^{2}$ is obtained:

\eqn\actioncuatro{
S_{t}^{(p+1)}=\int_{{\Sigma}_{p+1}}d^{p-1} \widetilde{x} d^{2}w\widetilde{V_{*}}(\overline{T},T).}

Notice that in the case when the gauge field and the tachyon $*$-commute,
$\widetilde{V_*}(\bar{T},T)$
reduces 
to a polynomial in $\bar{T} * T$. This is equivalent to assuming that the
gauge fields are independent of the noncommutative coordinates, from which we can deduce that 

\eqn\rlongdos{
R^{\mu }*\overline{T}*R_{\mu }*T=R^{\mu }\overline{T}*R_{\mu }T.}

We will stick to this result in order to avoid complications, since we are interested on how
noncommutativity affects the tachyon, not the gauge fields. 

\vskip 1truecm
\subsec{The Complex GMS Soliton}

Now, we are ready to move on and find a nontrivial solution to the complex tachyon of the form \trdos.
Let's rewrite \actioncuatro\ as

\eqn\acuatro{
S_{t}^{(p+1) }=\int_{{\cal M}^{(p-2)+1}}d^{p-1} \widetilde{x} S_{t}^{( *) },}
where

\eqn\astar{
S_{t}^{( *) }=\int_{\IR^2_*} d^{2}w\widetilde{V_{*}}( \overline{T},T)} 
is the action along the noncommutative plane. 

The equations of motion in $\IR_{*}^{2}$ the above action yields
are

\eqn\eqm{
{\partial \widetilde{V_{*}}( \overline{T},T) \over \partial T}=0, \ \ \ \ \  \ \ \ \ {\partial
\widetilde{V_{*}}( \overline{T},T)  \over \partial \overline{T}}=0.}

We cannot use the same solution as in HKLM's real bosonic case because the tachyon
is now charged \refs{\hklm}. Notice that, in the case of $\Theta =0$, the solutions
would simply solve to the following algebraic equations:

$$
{\partial \widetilde{V}( \overline{t},t) \over \partial T}=R^{\mu }\overline{t}R_{\mu
}+\sum ^{n}_{k=2}ka_{k}( \overline{t}t) ^{k-1}\overline{t}=0,
$$

\eqn\algebraic{
{\partial \widetilde{V}( \overline{t},t) \over \partial \overline{t}}=R^{\mu }R_{\mu }t+\sum
^{n}_{k=2}ka_{k}t( \overline{t}t) ^{k-1}=0,}
where $t$ is a scalar complex field. Such solutions are just constants in the commutative plane,
$\IR^{2}$. 

We know that the introduction of noncommutativity allows more interesting solutions \doce.
From \eqm, we see that the equations of motion in $\IR_{*}^{2}$
are

$$
{\partial \widetilde{V_{*}}( \overline{T},T) \over \partial T}=R^{\mu
}\overline{T}R_{\mu }+\sum ^{n}_{k=2}ka_{k}( \overline{T}*T) ^{k-1}*\overline{T}=0,
$$

\eqn\nonalg{
{\partial \widetilde{V_{*}}( \overline{T},T) \over \partial \overline{T}}=R^{\mu
}R_{\mu }T+\sum ^{n}_{k=2}ka_{k}T*( \overline{T}*T) ^{k-1}=0.}

Let's construct a simple complex solution of the form

\eqn\sol{
T=t_{*}T_{0}, \ \ \ \ \ \ \ \
\overline{T}=\overline{t_{*}}\overline{T_{0}},}
where $ T_{0} $ and $ \overline{T_{0}} $ have the following property:

\eqn\restr{
( \overline{T_{0}}*T_{0}) *( \overline{T_{0}}*T_{0}) =(
\overline{T_{0}}*T_{0}).}

In the commutative case, we would not be able to construct a nontrivial function $\bar{T}_0 * T_0$ that
squares to itself. This is just one of the many amazing properties the $*$-product has. Let's now see what
happens when we insert solution \sol\ into the equations of motion \nonalg:

$$
R^{\mu }(\overline{t_{*}}\overline{T_{0}})R_{\mu }+\sum ^{n}_{k=2}ka_{k}\bigg(
(\overline{t_{*}}\overline{T_{0}})*(t_{*}T_{0})\bigg)
^{k-1}*(\overline{t_{*}}\overline{T_{0}})=0,
$$

\eqn\plug{
R^{\mu }R_{\mu }(t_{*}T_{0})+\sum
^{n}_{k=2}ka_{k}(t_{*}T_{0})* \bigg( (\overline{t_{*}}\overline{T_{0}})*(t_{*}T_{0})\bigg)
^{k-1}=0.}

Now, since $t_{*}$ and $ \overline{t_{*}}$ are simply constants under
the $*$-product, they can be carried around the equations. Therefore, Eq. \plug\
can be rewritten as

$$
R^{\mu }\overline{t_{*}}R_{\mu }\overline{T_{0}}+\sum ^{n}_{k=2}ka_k(
\overline{t_{*}}t_{*}) ^{k-1} ( \overline{T_{0}}*T_{0})
^{k-1} * (\overline{t_{*}}\overline{T_{0}})=0,
$$

\eqn\plugdos{
T_{0}R^{\mu }R_{\mu }t_{*}+\sum
^{n}_{k=2}ka_{k}t_{*}( \overline{t_{*}}t_{*}) ^{k-1} T_0 *(
\overline{T_{0}}*T_{0}) ^{k-1}=0.}

Next, ${*}$-multiply the first equation by $T_{0}$ on the right, and the
second by $\overline{T_{0}}$ on the left, thereby obtaining

$$
R^{\mu }\overline{t_{*}}R_{\mu }\overline{T_{0}}*T_{0}+\sum
^{n}_{k=2}k a_k ( \overline{t_{*}}t_{*}) ^{k-1}( \overline{T_{0}}*T_{0})
^{k}\overline{t_{*}}=0,
$$

\eqn\plugdoscinco{
\overline{T_{0}}*T_{0}R^{\mu }R_{\mu }t_{*}+\sum
^{n}_{k=2}ka_{k}t_{*}( \overline{t_{*}}t_{*}) ^{k-1}(
\overline{T_{0}}*T_{0}) ^{k}=0.}

Notice that from the property  \restr\ we can deduce by iteration that

\eqn\restrdos{
( \overline{T_{0}}*T_{0}) ^{k}=( \overline{T_{0}}*T_{0}),}
where $k$ is a positive integer (i.e. the term $\bar{T}_0 * T_0$ behaves like a projection operator).

Therefore, using the above result, we can rewrite \plugdos\ as
$$
\bigg( R^{\mu }\overline{t_{*}}R_{\mu }+\sum ^{n}_{k=2} k a_k ( \overline{t_{*}}t_{*})
^{k-1}\overline{t_{*}}\bigg) \overline{T_{0}}*T_{0}=0,
$$

\eqn\plugtres{
\overline{T_{0}}*T_{0}\bigg( R^{\mu
}R_{\mu }t_{*}+\sum ^{n}_{k=2}ka_{k}t_{*}( \overline{t_{*}}t_{*}) ^{k-1}\bigg) =0.}
Since we are searching nontrivial solutions, we know that when $\bar{T}_0 * T_0 \not= 0$, the following
equations hold:

$$
R^{\mu }\overline{t_{*}}R_{\mu }+\sum ^{n}_{k=2}ka_{k}( \overline{t_{*}}t_{*})
^{k-1}\overline{t_{*}}=0, \ \ \ \ \ \
$$

\eqn\algebraicdos{
R^{\mu }R_{\mu }t_{*}+\sum ^{n}_{k=2}ka_{k}t_{*}(
\overline{t_{*}}t_{*}) ^{k-1}=0.}
These are precisely the algebraic equations of motion \algebraic\ in
the case of absent noncommutativity, with $t=t_{*}$ and $\overline{t}=\overline{t_{*}}$. 

In summary, we found that the coefficients $t_*$ and $\bar{t_*}$ of the solution we have constructed are
themselves solutions to the algebraic (commutative) equations:

\eqn\eqmdos{
{\partial \widetilde{V}( \overline{t},t) \over \partial t}=0,  \ \ \ \ \ \ \  {\partial
\widetilde{V}( \overline{t},t) \over \partial \overline{t}}=0.}

Our task now is to find $T_{0}$ and $\overline{T_{0}}$ such that they satisfy $
(\overline{T_{0}}*T_{0}) *( \overline{T_{0}}*T_{0}) =(
\overline{T_{0}}*T_{0}).$ Notice that, via the WWM correspondence, we can associate the fields $T_{0}$ and
$\overline{T_{0}}$ to the operators 

\eqn\proj{
\widehat{T_{0}}=i\widehat{P}, \ \ \ \ \ \ \ 
\widehat{\overline{T_{0}}}=-i\widehat{P},}
in $ {\cal H}( \widehat{a},\widehat{a}^{\dagger })$, where
$\widehat{P}$ is the projection operator $\widehat{P}=\widehat{P}^2.$ In the SHO basis any
projection operator may be expressed as  $\widehat{P}=\left| n\right\rangle
\left\langle n\right|.$

According to the WWM correspondence, the operator $\left| n\right\rangle \left\langle n\right|$ in ${\cal
H}(\widehat{a},\widehat{a}^{\dagger })$ is related in to the Wigner function $
2(-1)^{n}e^{-r^{2}}L_{n}(2r^{2})$ in $\IR_{*}^{2}$, where $L_{n}\left( s\right)$ is a Laguerre polynomial
\refs{\doce}. It can be shown that the general solution is a linear combination of projection operators
(i.e. Wigner functions) with complex coefficients that minimize the commutative potential
$\widetilde{V}(t,\bar{t})$. However, for the time being, we will only focus on the state in the lowest
energy which is given by the Gaussian packet $T_{0}(r^{2})=2e^{-r^{2}}$, where
$r^{2}=x^{2}+y^{2}=w\overline{w}+\overline{w}w$ and $L_{0}(s)=1$.

Summarizing, from the complex tachyon in ${\cal M}^{(p-2)+1}\otimes \IR_{*}^{2}$
we have a complex GMS soliton of the form:

\eqn\ncs{
T(w,\overline{w})=2it_{*}e^{-r^{2}}, \ \ \ \ \ \ 
\overline{T}\left( w,\overline{w}\right) =-2i\overline{t_{*}}e^{-r^{2}},}
where $t_{*}$ and $\overline{t_{*}}$ minimize the algebraic equation

\eqn\ap{
\widetilde{V}\left( t_*,\overline{t}_*\right) =0.}

It is remarkable that the only information
we need to know about the potential $\widetilde{V}$ is that it possesses
at least two local minima, and the values of $T$ and $\overline{T}$ for which these
would be minimized if noncommutativity is absent. 

This object may be interpreted as the low energy remnant of a stable $D(p-2)$-brane
arising from the annihilation of the unstable non-BPS $Dp-\overline{Dp}$-brane
pair in Type II theory.

In the case $p=3$ in Type IIB theory, we coin the term noncommutative string for the
resulting complex GMS soliton ({\it which is itself the low energy remnant of the
$D$-string}).

In the following section, we are going to apply all the $\IR_{*}^{2} \leftrightarrow {\cal
H}(\widehat{p},\widehat{q})$
technology to obtain an effective theory along the noncommutative string
from a theory with left-handed fermions in ${\cal M}^{1+1}\otimes \IR^{2}_{*},$ and show that
the conductivity on this object persists.

\vskip 2truecm
\newsec{ The Noncommutative String in the Presence of Fermions}

In this second part of this work, we will show the existence of an analog to Witten's
original
superconducting string \trece\ in the context of noncommutative solitons and $D$-brane annihilations
in
string theory. 

The idea is to begin with a $D3-\bar{D3}$-brane configuration in Type IIB superstring theory and in the
presence of a large background $B$-field turned on along the $x-y$ plane in the worldvolume. Such system
is unstable
and decays into a $D1$-brane, which has our complex GMS soliton as its low-energy remnant. The open string
attached to the $D1$-brane has chiral fermions in its supersymmetric spectrum. This is because this
spectrum is induced from the ten-dimensional Type IIB theory, which is chiral. Such fermions see the
complex noncommutative soliton has a background field. By applying the WWM correspondence, we will
integrate out the noncommutative coordinates and find an effective two-dimensional worldsheet theory along
our $D$-string. In Sec. 4.1, for the sake of simplicity, we will first integrate the case where the
gauge fields $R_{\mu}$ are absent. In Sec. 4.2 we introduce the gauge fields $R_{\mu}$, which appear as a
``mass'' term in the effective theory. The bosonization technique is used in section 4.3 to display
superconductivity.

\vskip 1truecm
\subsec{Free Fermions in ${\cal M}^{1+1}\otimes \IR_{*}^{2}$}

In ${\cal M}^{1+1} \otimes \IR^2_*$ we can express left-handed Dirac spinors as

\eqn\cuatrospinor{
\Psi = \pmatrix{0 \cr
\psi_{L}\cr},}
where $\psi _{L}$ is a two-component spinor obeying the Weyl equation 
$\overrightarrow{\sigma }\cdot \widehat{p}\psi _{L}=-\psi _{L}.$
In the above equations, $\widehat{p} = {\overrightarrow{p} \over \left| \overrightarrow{p}\right| }$,
where $\overrightarrow{p}$ is the spatial part of the fermion's momentum. Also, 
$\overrightarrow{\sigma }=( \sigma ^{1},\sigma ^{2},\sigma ^{3})$, where $\sigma ^{i}$ are the
well-known Pauli matrices. Thus,

\eqn\pauli{
\overrightarrow{\sigma }\cdot \overrightarrow{p}= \pmatrix{
p_3 & p_1-ip_2\cr
p_1+ip_2 & -p_3}.}

In four dimensions, the free fermions satisfy the massless Dirac equation

\eqn\dirac{
i\partial \! \! \! /\Psi =0,}
where $\partial \! \! \! /=\gamma ^{\mu }\partial _{\mu }$, and  $\gamma ^{i}= \pmatrix{
0 & -\sigma ^i\cr
\sigma^i & 0}$, $ \gamma ^{0}= \pmatrix{
0 & \sigma^0\cr
\sigma^0 & 0},$
are the Dirac matrices and where $\sigma^0$ is a $2 \times 2$ unit
matrix\foot{Notation: In this section, we will denote the indices $\mu,\nu, \dots = 0,1,2,3$;
$i,j,
\dots = 1,2,3$; $a,b, \dots = 0,3$ (commuting coordinates) and $\alpha, \beta, \dots = 1,2$
(noncommuting
coordinates).}. These matrices satisfy the Clifford algebra $\{ \gamma ^{\mu },\gamma ^{\nu } \}
=2\eta
^{\mu \nu}.$

Define the chirality operator $\gamma^{5}=i\gamma ^{0}\gamma ^{1}\gamma ^{2}\gamma ^{3}= \pmatrix{
\sigma^0 & 0\cr
0 & - \sigma^0}.$ We can define a left-handed spinor $\Psi_{L}={1 \over 2}( \widehat{1}-\gamma
^{5}) \Psi =
\pmatrix{0 \cr
\psi_{L}\cr},$
where $\widehat{1}$ is the $4\times 4$ unit matrix and $\psi_L$ obeys the chirality
equation: $\gamma ^{5}\psi _{L}=-\psi _{L}.$ With this background, we now are ready to introduce the
noncommutative string
defined in \ncs. 

The action for fermions in the presence of this object
has the following generic form of Yukawa couplings \refs{\vcuatro}

\eqn\scuatrof{
S_{f}^{( 4) }=\int dtdzd^{2}w \bigg( f( \overline{T}) *\overline{\Psi }*g( T) *\gamma ^{\mu }\partial _{\mu
}\Psi\bigg) ,}
where $f$ and $g$ are polynomials similar to \poly, which play the role of fermion-soliton coupling. 
Therefore, using \restr, we find that

\eqn\efege{
f(\overline{T}) =f( \overline{t_{*}}) \overline{T_{0}}, \ \ \ \ \ \ \ \ 
g(T) =g(t_{*}) T_{0}.}

Now, we know $\overline{\Psi }=\Psi^{\dagger}\gamma^{0}= \pmatrix{0\cr 
\psi _{L}\cr}^{\dagger} \pmatrix{0 & \sigma^0 \cr
\sigma^0 & 0\cr}= (\overline{\psi _{L}}, 0).$
Thus, the action \scuatrof\ can be reexpressed as

\eqn\scuatrofdos{
S^{( 4) }_{f}=\int dtdzd^{2}wf( \overline{t_{*}}) g( t_{*}) \bigg( \overline{T_{0}}*(
\overline{\psi _{L}}, 0) *T_{0}*\gamma ^{\mu }\partial _{\mu } \pmatrix{
0\cr
\psi _{L}\cr} \bigg) .}
In rescaled units of the noncommutativity parameter  $\Theta$, Dirac operator is written

\eqn\gammadelta{
\gamma ^{\mu }\partial _{\mu }=\gamma ^{a}\partial _{a}-{1 \over \sqrt{\Theta
}} \gamma ^{\alpha}\partial _{\alpha}.}
In the limit $\Theta \to \infty$, we get

\eqn\gammadeltados{
\gamma^{\mu }\partial_{\mu }=\gamma^{a}\partial_{a},}
and
$$
S^{( 4) }_{f}=\int dtdzd^{2}wf( \overline{t_{*}}) g( t_{*}) \bigg[ \overline{T_{0}}*(\psi _{L}, 0) 
*T_{0}*( \gamma ^{0}\partial _{0}-\gamma ^{3}\partial _{3}) \pmatrix{0\cr
\psi _{L}\cr} \bigg].
$$

Applying the WWM correspondence and recalling the trace formula \trace, let's rewrite the action above
as

\eqn\trf{
S^{( 4) }_{f}=2\pi \Theta f( \overline{t_{*}}) g( t_{*}) \int dtdzS_{f}^{( *) },}
where the action along
the noncommutative coordinate plane (written in terms of the two-component
spinors) is

$$ 
S_{f}^{( *) }=Tr\left\{ \widehat{\overline{T_{0}}}(\widehat{\overline{\psi _{L}}},0) \widehat{T_{0}} \big[
\pmatrix{0 & \sigma ^{0}\cr
\sigma ^{0} & 0\cr} \pmatrix{0\cr
\partial _{0}\widehat{\psi _{L}}\cr} +  \pmatrix{0 & \sigma ^{3}\cr
-\sigma ^{3} & 0\cr} \pmatrix{0\cr
\partial _{3}\widehat{\psi _{L}}\cr}\big] \right\}  
$$

\eqn\trfstar{
=Tr \bigg[ (\widehat{\overline{T_{0}}}\, \widehat{\overline{\psi _{L}}},
\widehat{\overline{T_{0}}}\,
\widehat{\overline{\psi _{R}}}) \pmatrix{\widehat{T_{0}}\sigma ^{0}\partial _{0}\widehat{\psi
_{L}}+\widehat{T_{0}}\sigma ^{3}\partial
_{3}\widehat{\psi _{L}}\cr
0\cr} \bigg].}
This may be rewritten as
\eqn\sfasterisco{
S_{f}^{( *) }=Tr \bigg( \widehat{\overline{T_{0}}}\, \widehat{\overline{\psi
_{L}}}\widehat{T_{0}}\sigma
^{0}\partial _{0}\widehat{\psi _{L}}+\widehat{\overline{T_{0}}}\, \widehat{\overline{\psi
_{L}}}\widehat{T_{0}}\sigma ^{3}\partial _{3}\widehat{\psi _{L}} \bigg).}

Now, using \expand\ and the WWM correspondence, in the SHO basis, we expand

\eqn\expdos{
\widehat{\psi }_{L}( x^{\mu }) =\sum _{m,n\geq 0}\psi ^{L}_{mn}( z,t) \left| m\right\rangle
\left\langle n\right| .}

Indeed, having obtained this:

\eqn\projcuatro{
\widehat{T_{0}}=i\left| 0\right\rangle \left\langle 0\right| , \ \ \ \ \ \
\widehat{\overline{T_{0}}}=-i\left| 0\right\rangle \left\langle 0\right| ,}
we are in the position to calculate the trace of a generic term of the form 
$\widehat{\overline{T_{0}}}\widehat  {\overline{\psi_L}}\widehat{T_{0}}D\widehat{\psi_L}$, where $D$
is a
$2\times 2$ matrix differential operator. Thus,

$$
Tr\bigg( \widehat{\overline{T_{0}}}\, \widehat{\overline{\psi_L}}\widehat{T_{0}}D\widehat{\psi_L}
\bigg)
=Tr \bigg[
(-i\left| 0\right\rangle \left\langle 0\right| )(\sum _{m,n\geq 0}\overline{\psi_{mn}^L}\left|
m\right\rangle \left\langle n\right| )(i\left| 0\right\rangle \left\langle 0\right| )D(\sum _{r,s\geq
0}\psi_{rs}^L\left| r\right\rangle \left\langle s\right| ) \bigg]
$$

\eqn\generic{
=Tr \bigg( \left| 0\right\rangle \sum _{m,n,r,s\geq 0}( \overline{\psi_{mn}^L}D\psi_{rs}^L) (
\left\langle 0\mid
m\right\rangle \left\langle n\mid 0\right\rangle \left\langle 0\mid r\right\rangle \left\langle s\right|
) \bigg) .}
In the process, we have used the fact that the kets $\left| n\right\rangle$ form a complete orthonormal
basis which,
by definition, satisfy $\left\langle m\mid n\right\rangle =\delta _{mn}.$
Also, each ket $\left| n\right\rangle$ is applied into a one-dimensional
subspace of the Hilbert space. This means that $Tr_{\cal H}(\left| m\right\rangle \left\langle n\right|
)=\delta _{mn}.$ Applying these facts, we deduce that

$$
Tr \bigg(\widehat{\overline{T_{0}}}\, \widehat{\overline{\psi}_L}\widehat{T_{0}}D\widehat{\psi_L}
\bigg)
=\sum
_{m,n,r,s\geq 0} \bigg( ( \overline{\psi^L_{mn}}D\psi^L_{rs}\delta _{0m}\delta _{n0}\delta _{0r}) Tr(
\left|
0\right\rangle \left\langle s\right| ) \bigg)
$$

\eqn\cerocero{
=\overline{\psi^L_{00}}D\sum _{s\geq 0}\psi^L_{0s}\delta _{0s}=\overline{\psi^L_{00}}D\psi^L_{00}.}
With this result, the action on the noncommutative plane is

\eqn\sncm{
S_{f}^{( *) }=\overline{\psi ^{L}_{00}}\sigma ^{0}\partial _{0}\psi ^{L}_{00}+\overline{\psi
^{L}_{00}}\sigma ^{3}\partial _{3}\psi ^{L}_{00}.}

In the performing of the trace, we actually integrated out $w$ and $\overline{w}$.
Also, notice how the properties of the projection operators $T_0$ and $\bar{T}_0$ have ``projected out''
most of the $\psi_{mn}^L(z,t)'s$, leaving behind just the $\psi_{00}^L(z,t)$ term in the effective
two-dimensional theory along the noncommutative $D$-string. Therefore, the left-handed fermionic action
along the noncommutative string is

\eqn\ncsfa{
S^{( ncs) }_{f} = S_f^{(4)} = 2\pi \Theta f( \overline{t_{*}}) g( t_{*}) \int dtdz (\overline{\psi
^{L}_{00}}\sigma
^{0}\partial _{0}\psi ^{L}_{00}+\overline{\psi ^{L}_{00}}\sigma ^{3}\partial _{3}\psi
^{L}_{00}).}
This is precisely the localization of chiral fermions on the
$D1$-string, done with the techniques utilized in Ref. \refs{\vcuatro}. In the present case the chiral
fermions are localized on the 
noncommutative $D$-string.

From this point on, we shall avoid the use of unnecessary subindices, since these yield
no information when the effective theory on ${\cal M}^{1+1}$ is studied.
Thus, we will simply use

\eqn\sub{
\psi ^{L}_{00}( z,t) =\psi ^{L}( z,t). }

Thus, the effective action for left-handed fermions along the string is

\eqn\ncst{
S^{( ncs) }_{f}=2\pi \Theta f( \overline{t_{*}}) g( t_{*}) \int dtdz \big( \overline{\psi }^L
\sigma^{a}\partial_{a}\psi^L \big).}

It is time to move on and generalize this result to the case when gauge fields
are turned on.

\vskip 1truecm
\subsec{$U(1)\otimes \widetilde{U}(1)$ Gauge-coupled Fermions
in ${\cal M}^{1+1}\otimes \IR_{*}^{2}$}

The appearance of gauge fields arising from the Chan-Paton factors should
supply with further properties characterizing the noncommutative string. The introduction of this gauge
field merely amounts, as usual, to minimally coupling fermions
to $R^{\mu}$: $\partial _{\mu }\Psi \to D_{\mu }\Psi =(\partial _{\mu }-iR_{\mu })\Psi .$

So, the four-dimensional action for gauge-coupled fermions in the presence of
the noncommutative string is

\eqn\scuatrofg{
S_{gauge}^{(ncs) }=\int dtdzd^{2}w \bigg( f(\overline{T})*\overline{\Psi }*g(T)D\! \! \! \! /\Psi
\bigg) =\int
dtdzd^{2}w \bigg( f(\overline{T})*\overline{\Psi }*g(T)*\gamma ^{\mu }(\partial _{\mu }\Psi -iR_{\mu }*\Psi )\bigg)
,}
which may be written as

\eqn\fplusg{
S^{(ncs) }_{gauge}=S_{f1}^{(ncs)}+ S_{f2}^{(ncs) },}
where $S_{f1}^{(ncs)}$ is given by \scuatrof\ and

\eqn\scuatrog{
S^{(ncs)}_{f2}=-i\int dtdzd^{2}w \bigg( \gamma ^{\mu }R_{\mu }f(\overline{T})*\overline{\Psi
}*g(T)*\Psi \bigg)} 
is the contribution due to the presence of the gauge fields\foot{
Unlike \expdos, there is no need to expand $R_{\mu }$, because it
is constant on the noncommutative plane. This condition is equivalent to saying that $R_{\mu}$ and the
tachyon commute (see Eq. \rlongdos).}.

Applying the WWM correspondence, the action \scuatrog\ is written as

$$
S_{f2}^{(ncs) }=-2\pi i\Theta f( \overline{t}_{*}) g( t_{*}) \int dtdzTr\bigg(
\widehat{\overline{T_{0}}}\,
\widehat{\overline{\Psi }}\widehat{T_{0}}\gamma ^{\mu }R_{\mu }\widehat{\Psi }\bigg) 
$$
\eqn\scuatrogdos{
=-2\pi \Theta if(
\overline{t_{*}}) g( t_{*}) \int dtdzS_{f2}^{( *) },}
where

\eqn\sgstar{
S_{f2}^{(*)}=Tr \bigg( \widehat{\overline{T_{0}}}\, \widehat{\overline{\Psi }}\widehat{T_{0}}\gamma
^{\mu
}R_{\mu }\widehat{\Psi } \bigg)}
is the gauge-field contribution to the action on the noncommutative plane.

Using the relation

\eqn\gammar{
\gamma ^{\mu }R_{\mu }= \pmatrix{0 & \sigma ^{0}R_{0}\cr
\sigma ^{0} R_{0} & 0\cr}  - \pmatrix{0 & \sigma ^{i}R_{i}\cr
-\sigma ^{i}R_{i} & 0\cr},}
we rewrite \sgstar\ as
$$
S_{f2}^{( *) }=Tr\left\{ \widehat{\overline{T_{0}}}( \widehat{\overline{\psi
_{L}}},0) \bigg[\pmatrix{0 & \widehat{T_{0}}\sigma ^{0}R_{0}\cr
\widehat{T_{0}}\sigma ^{0}R_{0} & 0\cr} - \pmatrix{0 & \widehat{T_{0}}\sigma ^{i}R_{i}\cr
-\widehat{T_{0}}\sigma ^{i}R_{i} & 0\cr} \bigg] \pmatrix{0\cr
\widehat{\psi _{L}}\cr} \right\}
$$

\eqn\sgrcuatrogr{
=Tr \bigg[ (\widehat{\overline{T_{0}}}\, \widehat{\overline{\psi _{L}}},
\widehat{\overline{T_{0}}}\,0) \pmatrix{ \widehat{T_{0}}\sigma
^{0}R_{0}\widehat{\psi _{L}}-\widehat{T_{0}}\sigma ^{i}R_{i}\widehat{\psi _{L}}\cr
0\cr} \bigg],}
or equivalently as

\eqn\sgstarexp{
S_{f2}^{( *) }=Tr \bigg( \widehat{\overline{T_{0}}}\, \widehat{\overline{\psi
_{L}}}\widehat{T_{0}}\sigma
^{0}R_{0}\widehat{\psi _{L}}-\widehat{\overline{T_{0}}}\, \widehat{\overline{\psi
_{L}}}\widehat{T_{0}}\sigma ^{i}R_{i}\widehat{\psi _{L}} \bigg).}

Recalling \expdos\ and \proj, let's now calculate the trace of a generic term of the form
$\widehat{\overline{T_{0}}}\, \widehat{\overline{\psi}_L}\widehat{T_{0}}R\widehat{\psi}_L$,
where $R$ is a $2\times 2$ matrix field independent of $w$ and
$\overline{w}$. 

After some computations similar to those of the previous section, we are left with

$$ 
Tr\bigg( \widehat{\overline{T_{0}}}\, \widehat{\overline{\psi}_L}\widehat{T_{0}}R\widehat{\psi}_L\bigg) 
=\overline{\psi^L_{00}}R\psi^L_{00.}
$$

Therefore, the gauge field contribution to the action along the noncommutative
plane is

\eqn\sncg{
S_{f2}^{( *) }=\overline{\psi ^{L}_{00}}\sigma ^{0}R_{0}\psi ^{L}_{00}-\overline{\psi
^{L}_{00}}\sigma
^{i}R_{i}\psi ^{L}_{00}.} 

Having integrated out the coordinates $w$ and $\overline{w}$, the
gauge-field contribution to the action along the noncommutative string is

\eqn\ncsg{
S^{(ncs) }_{f2}=-2\pi i \Theta f( \overline{t_{*}}) g( t_{*}) \int dtdz (\overline{\psi
^{L}_{00}}\sigma
^{0}R_{0}\psi ^{L}_{00}-\overline{\psi ^{L}_{00}}\sigma ^{i}R_{i}\psi ^{L}_{00}).}

Getting rid of unnecessary subindices, we have 

\eqn\ncsgdos{
S^{(ncs)}_{f2}=-2\pi i\Theta f( \overline{t_{*}}) g( t_{*}) \int dtdz \big[ \overline{\psi }^L\sigma
^{\mu}R_{\mu}\psi^L \big]. }
Since $\sigma^{\mu} R_{\mu} = \sigma^a R_a - \sigma^{\alpha}R_{\alpha}$ and we may define the
matrix mass parameter
$ m(z,t) \equiv \sigma^{\alpha}R_{\alpha}.$ Therefore, Eq. \ncsgdos\ can be expressed as 

\eqn\ncsfg{
S^{( ncs) }_{f2}=-2\pi \Theta if( \overline{t_{*}}) g( t_{*}) \int dtdz \bigg( i\overline{\psi
}^L\sigma
^{a} R_{a}\psi^L - \overline{\psi }^L m \psi^L \bigg).}
In conclusion, the complete gauged-coupled action along the string is:

\eqn\ncsfgdos{
S^{( ncs) }_{gauge}=-2\pi \Theta if( \overline{t_{*}}) g( t_{*}) \int dtdz \bigg( i\overline{\psi
}^L\sigma
^{a} D_{a}\psi^L - \overline{\psi }^L m \  \psi^L \bigg),}
where $D_a \psi_L(z,t) = (\partial_a -i R_a) \psi_L(z,t).$

With all these tools, we are ready to calculate the current along this object.

\vskip 1truecm
\subsec{The Bosonization Technique: Superconductivity}

For the time being, we will focus in massless case; {\it i.e.}, $R_1 = R_2 = 0$ and $R_a \not= R_a(z),$
$a=0,3.$
First of all, let's rescale the action of gauge-coupled fermions along the noncommutative
string such that the coefficient outside the integral is set equals to one.
So, upon reintroducing the gauge potential

\eqn\faction{
S^{(ncs)}_{gauge} = \int dzdt \bigg( i\overline{\psi_L }\sigma ^{a}D_{a}\psi_L \bigg) .}

In any theory with fermions in two dimensions, we can equivalently use bosons or fermions by
applying the technique of bosonization. The idea is to introduce a scalar field 
$\zeta (z,t)$ living on the noncommutative string:

\eqn\bosonization{
\overline{\psi_L }\sigma ^{a}\psi_L ={1 \over \sqrt{\pi }}\varepsilon ^{ab}\partial _{b}\zeta.}

Thus according to Ref. \refs{\trece}, a two-dimensional kinetic term is

\eqn\fb{
i\overline{\psi_L }\sigma ^{a}D_{a}\psi_L = {1 \over 2} (\partial_a \zeta) (\partial^a \zeta) -
{1 \over \sqrt{\pi}} R_a \varepsilon^{ab} \partial_b \zeta,}
which yield the following equation of motion:

\eqn\parcial{
\partial_a \partial^a \zeta + 
{1 \over \sqrt{\pi}} E  = 0,}
where $E=\varepsilon^{ab} \partial_a R_b$ is the electric field in two dimensions.
 
Now, the conserved current is just $J^a = - \overline{\psi_L} \sigma^a \psi_L$, which means that from
Eq. (4.34) that $J^3 = -{1 \over \sqrt{\pi}} \ddot{\zeta}.$ From (4.36) and the $z$-independence of $R_a$
we see that $\ddot{\zeta} =  -{1 \over \sqrt{\pi}} E.$ Thus, we get for $J^3$ (the current along the
string) that

\eqn\superconductivity{
{dJ^3(z,t) \over dt}={1 \over \pi }E.}
This equation means that the string is superconducting. If an electric field $E$ is applied for some
time $T$ a current ${ET \over \pi}$ remains even is the electric field is turned off after time $T$. 

For a regular wire of finite conductivity $\sigma$, the current is $J^3=\sigma E^3$
(where $E^3$ the component of the electric field along the string)
and vanishes after a certain characteristic time if $E^3$ is turned off.
The situation for our noncommutative string is quite similar to the Witten's superconducting string.

Conservation of the fermionic current could be related to the conservation of some fermionic numbers, such
as the lepton and baryon numbers of the theory on the brane. It would be very interesting to construct
specific brane configurations of intersecting branes which reproduces Standard Model and some GUT's with
superconducting noncommutative $D$-strings. Here the fermionic conserved current will be directly related
to the fermionic quantum numbers of baryon and lepton numbers of the underlying reproduced models. This
will be reported in a forthcoming communication.

\vskip 2truecm
\newsec{Final Remarks}

In the present paper, $D$-brane annihilation and noncommutativity were merged
together to obtain a new object: a noncommutative string with nondecaying conductivity.

The necessary constituents to construct this entity were all present in the Type IIB superstring
theory. By
rotating one
of them an angle $\pi$ in the transverse directions, we turned it into a $\overline{Dp}$-brane.
The result was a non-BPS $Dp-\overline{Dp}$-brane system, which is unstable due to the presence
of a tachyon in its worldvolume. On the other hand, the NS-NS sector gave
rise to the ubiquitous background $B$-field, which played a pivotal role in the introduction of
noncommutativity.

The predominant approach to such an annihilation has been to find a vortex-like configuration of the
tachyon field, thereby obtaining a stable BPS $D(p-2)$-brane as the result. The tachyon in the
$Dp-\overline{Dp}$-brane worldvolume is charged under the gauge group
$U(1)\otimes \widetilde{U}(1)$ arising from the Chan-Paton factors on each $D$-brane. Assuming
we
have a flat metric, we introduce a constant $B$-field along two spatial directions. In the low-energy
limit one obtain an effective noncommutative
theory where the fields are Moyal $*$-multiplied. From here on, we generalize the work of GMS to
the case where the field is complex and gauge-coupled.

For definiteness, we only discuss potentials of polynomial form (3.3) Also, since we are only interested
in how noncommutativity acts on the tachyon, we assume that it does not affect the gauge fields and that
Eq. (3.10) is satisfied. This merely amounts to redefining the potential to $\widetilde{V}(T,\bar{T}) =
R^{\mu}T*R_{\mu}T$, which is itself also a polynomial in $\bar{T}*T.$

With this result, we show the natural existence of an object
analogous of Witten's
superconducting string \trece, in the context of noncommutative soliton theory. By making use of the
WWM correspondence, we find that the noncommutative $D$-string in the large noncommutativity limit
($\Theta \to \infty$) is completely specified by Eq. (3.26)

Starting with Type IIB theory $D3-\overline{D3}$ annihilation with a $B$-field turned on along the $x-y$
plane, the complex GMS is the remnant of a BPS $D$-string. From the localization of the chiral fermions
$\psi_L$ in the supersymmetric spectrum of the open sector (in the sense of Ref. \refs{\vcuatro}), we may
construct a two-dimensional effective description of the fermionic degrees of freedom along the
commutative coordinates $(z,t)$. This is done by integrating out the two noncommutative transverse
coordinates $(w,\overline{w})$ and exploring the soliton's projector properties. Although we could have
calculated the current directly, we used the bosonization technique for simplicity. The open string sector
allows fermionic states in the worldvolume ${\cal M}^{1+1} \otimes \IR^2_*$. We find that, by obtaining an
equation of the type (4.42), the conserved current is a persistent one.

Future subsequent work might include the use of the bosonization technique to explore more types of
phenomena, such as light scattering by the noncommutative $D$-string (see \refs{\trece}). Also we
are interested in the construction of intersecting brane configurations, thereby reproducing the Standard
Model and some GUT's containing noncommutative superconducting $D$-strings. In such cases, it might be
possible to identify the existence of
conserved fermionic superconducting current with that of conserved lepton and baryon quantum
numbers. Likewise, we could make some progress in including finite-$\Theta$ effects and generalizing
to the case when the gauge
fields get affected by noncommutativity. Another issue to be consider is to explore the stability of our
solution. Some of these issues are currently under investigation.

\vfill
\centerline{\bf Acknowledgements}

We are greatful to A. G\"uijosa, O. Loaiza-Brito and M. Przanowski, for useful
discussions. We thank R. Tatar for pointing out Ref. [26]. This work was supported in part by the CONACyT
grants No. 33951E and 30420E.

\vskip 2truecm

\listrefs

\end